# A Reproducible Reality-to-VR Pipeline for Ecologically Valid Aging-in-Place Research

Ibrahim Bilau[1], Stacie Smith[2], Abdurrahman Baru[1], Marwan Shagar[1], Brian Jones[3], Eunhwa Yang, PhD[1]

[1]School of Building Construction, Georgia Institute of Technology, Atlanta, Georgia

[2]School of Architecture, Georgia Institute of Technology, Atlanta, Georgia

[3]Institute for People and Technology, Georgia Institute of Technology, Atlanta, Georgia, USA

**Abstract**

Virtual reality (VR) has emerged as a promising tool for assessing instrumental activities of daily living (IADLs) in older adults. However, the ecological validity of these simulations is often compromised by simplified or low-fidelity environmental design that fails to elicit a genuine sense of presence. This paper documents a reproducible Reality-to-VR pipeline for creating a photorealistic environmental simulation to support a study on cognitive aging in place. The proposed workflow captured the as-built kitchen of the Aware Home building at Georgia Tech using Terrestrial Laser Scanning (TLS) for sub-millimeter geometric accuracy, followed by point cloud processing in Faro SCENE, geometric retopology in SketchUp, and integration into Unreal Engine 5 via Datasmith with Lumen global illumination for high visual fidelity. The pipeline achieved photorealistic rendering while maintaining a stable 90 Hz frame rate, a critical threshold for mitigating cybersickness in older populations. The environment also enables instantaneous manipulation of environmental variables, such as switching between closed cabinetry and open shelving, providing experimental flexibility impossible in physical settings. Participant validation with 17 older adults confirmed minimal cybersickness risk and preserved sensitivity to the experimental manipulation, supporting the pipeline's feasibility for aging-in-place research and establishing a benchmark for future comparative studies.

***Keywords:*** *Virtual Reality, Terrestrial Laser Scanning, Aging in Place, Environmental Design, Unreal Engine*

# 1. Introduction

## 1.1  Virtual Reality as a Methodological Tool in Behavioral Science

Virtual reality (VR) has evolved from a niche technology into a robust methodological instrument within the behavioral sciences (Parsons, 2015; Riva et al., 2022), with its adoption accelerating substantially across clinical and cognitive research domains over the subsequent decade (Rizzo et al., 2018; Vlake et al., 2023). By enabling the simulation of complex, real-world scenarios within a controlled laboratory setting, VR offers researchers a unique solution to the historical trade-off between experimental control and ecological validity (Blascovich et al., 2002; Rizzo & Koenig, 2017). Unlike traditional psychometric tests, which often rely on decontextualized or static stimuli, immersive virtual environments (IVEs) allow for the assessment of instrumental activities of daily living (IADLs), such as meal preparation, financial management, or spatial navigation, in a manner that predictively models real-world performance (Parsons et al., 2017). This approach is particularly relevant to Cognitive-Aging-in-Place research, where home design factors such as storage configuration and spatial accessibility substantially influence IADL performance among older adults with mild cognitive impairment (MCI) (Machry et al., 2025).

While VR has demonstrated the capacity to elicit physiological and behavioral responses comparable to real-world settings in general populations (Sanchez-Vives & Slater, 2005), achieving this level of immersion requires greater environmental fidelity in older adults, particularly those with MCI, who are more vulnerable to inconsistencies due to age-related changes in sensory processing and sensorimotor integration (Tuena et al., 2020).

## 1.2 Research Gap: Environmental Fidelity in Aging Research

The global demographic shift toward an aging population has driven increased use of virtual reality for the early detection of cognitive decline and for the delivery of non-pharmacological interventions for MCI and Alzheimer's disease and related dementia (ADRD) (Garcia-Betances et al., 2015; D'Cunha et al., 2019; Tuena et al., 2020). Despite this growth, a critical methodological gap remains regarding the design and development of the virtual environments themselves. Emerging literature indicates that older adults rely heavily on consistent and familiar environmental cues in VR to sustain immersion, cognitive engagement, and a sense of agency, and that noticeable inconsistencies in visual or behavioral elements, such as unnatural movement or model-like appearance, may disrupt the processes of immersion, cognitive engagement, and sense of agency (Lundstedt et al., 2023; Diwan et al., 2022; Tuena et al., 2020). Recent empirical work examining visual accessibility among older adults with MCI has demonstrated that open shelving can reduce cognitive load, improve motivation, and support more efficient task performance, underscoring the importance of accurately modeling such environmental features in VR studies (Bilau et al., 2025). Yet a review of recent studies reveals that most VR applications for aging research rely on synthetic environments constructed via manual 3D modeling (Skurla et al., 2022; Tarnanas et al., 2013; Yang et al., 2025). These idealized models frequently omit the ecological complexity, such as organic clutter, surface imperfections, and context-specific ambient lighting, that characterizes familiar domestic settings for older adults. While frameworks for VR software architecture exist, such as the Unity Experiment Framework (Brookes et al., 2020) or *BehaveFIT* (Wienrich et al., 2021), these primarily address interaction mechanics rather than environmental photorealism. Conversely, the Architecture, Engineering, and Construction (AEC) industry has established rigorous "Scan-to-BIM" workflows for creating high-fidelity building replicas using Terrestrial Laser Scanning (TLS) (Valero et al., 2022). To date, these high-fidelity engineering workflows remain largely siloed and have not yet been systematically adapted for behavioral science. Consequently, the field lacks a validated translation protocol for converting existing, ecologically valid physical spaces into interactive virtual environments suitable for aging-in-place behavioral research.

## 1.3 Study Context and Objectives

To bridge this methodological divide, this paper documents a "Reality-to-VR" pipeline developed for the second phase of a longitudinal study on "Cognitive Aging in Place." This broader study examines the impact of kitchen cabinet configurations, specifically open shelving versus closed cabinetry, on visual accessibility and functional independence in older adults, especially those living with MCI (Machry et al., 2025; Bilau et al., 2025). Previous findings indicate that open shelving can enhance visual accessibility and lower cognitive workload for older adults with MCI (Bilau et al., 2025). To ensure the experimental environment aligned with the participants' mental models of a functional home, the study reconstructed a high-fidelity photorealistic simulation of Aware Home, an existing residential research facility at the Georgia Institute of Technology (Georgia Tech), rather than relying on a synthetic or idealized model. The simulation supported two experimental conditions: Condition A (Closed Cabinetry) and Condition B (Open Shelving), both modeled within identical spatial footprints to isolate visual accessibility as the sole independent variable. A reproducible workflow was detailed that integrates TLS with the Faro

Focus system, point cloud processing in Faro SCENE, and geometric retopology in SketchUp to deploy a photorealistic simulation in Unreal Engine 5. This paper aims to provide behavioral researchers with a validated technical protocol for creating photorealistic environmental simulations. The translation from physical to virtual was documented to establish a reproducible benchmark for high-fidelity environmental simulation in aging research, enabling future comparative studies to empirically evaluate whether scan-based environments yield more transferable behavioral findings than synthetic alternatives.

## 2. Method

### 2.1 The Aware Home Building

The Aware Home building, located at Georgia Tech, served as the physical reference environment for this study. Commissioned in 1999, the Aware Home functions as a dedicated living laboratory (5,040 sq ft) designed to facilitate interdisciplinary research on ubiquitous computing and aging-in-place technologies (Kidd et al., 1999). Distinguished from standard clinical laboratories by its authentic residential architecture, the facility includes two identical floor plans that support comparative control studies in a naturalistic home setting. The first-floor kitchen (Figure 1) was targeted for digitization based on three primary methodological criteria. First, the kitchen was identified as a critical domain for IADLs, as meal preparation is a key indicator of functional independence in older adults (Lawton & Brody, 1969). Second, the L-shaped counter configuration provides a representative example of the spatial navigation challenges common in American homes. Third, the drop-ceiling fixture with indirect cove lighting creates a complex illumination scenario. This setup offers a rigorous test of the VR pipeline's ability to render High-Dynamic-Range (HDR) lighting conditions essential for visual accessibility (IES, 2020).

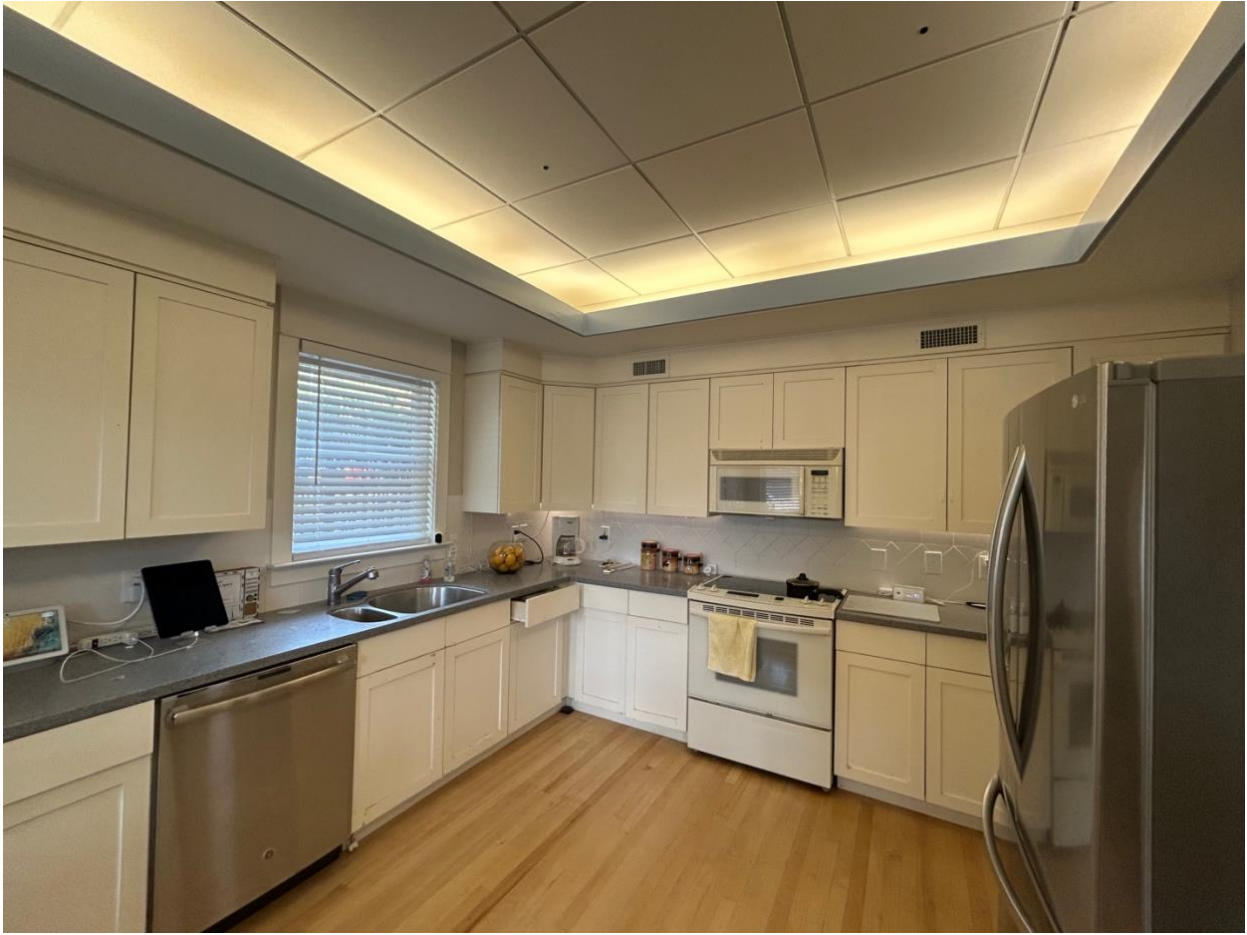

**Figure 1.** Physical kitchen setting in the Georgia Tech Aware Home. The L-shaped layout, drop-ceiling cove lighting, and reflective appliances presented challenges for achieving photorealistic digitization within the VR pipeline.

## 2.2 Terrestrial Laser Scanning

TLS, which employs Light Detection and Ranging (LiDAR) technology, provides superior accuracy for indoor mapping by capturing complex geometries and minimizing human measurement error (Becerik-Gerber et al., 2011). The space was scanned using a Faro Focus S70

TLS. To mitigate occlusion and data shadows caused by line-of-sight obstruction, a high-redundancy, multi-station scanning network was established. As illustrated in Figure 2, this protocol comprised two scan positions due to the kitchen's small size, and selected cabinet doors were detached to ensure comprehensive coverage of deep recesses behind cabinetry and beneath countertops. Prior to the scanning, a constellation of physical checkerboard registration targets was affixed to the vertical surfaces across the capture volume. These targets serve as a high-contrast reference for automated identification during post-processing. Deploying artificial targets is a standard best practice in surveying to maximize registration accuracy across overlapping datasets (Cheng et al., 2018). Furthermore, the scanner was configured to capture colorized data (RGB) alongside depth geometry. This enabled the generation of 360-degree HDR imagery at each station. These panoramic images were subsequently mapped onto the point cloud, providing a photorealistic reference layer critical for the texture and surface validation phase.

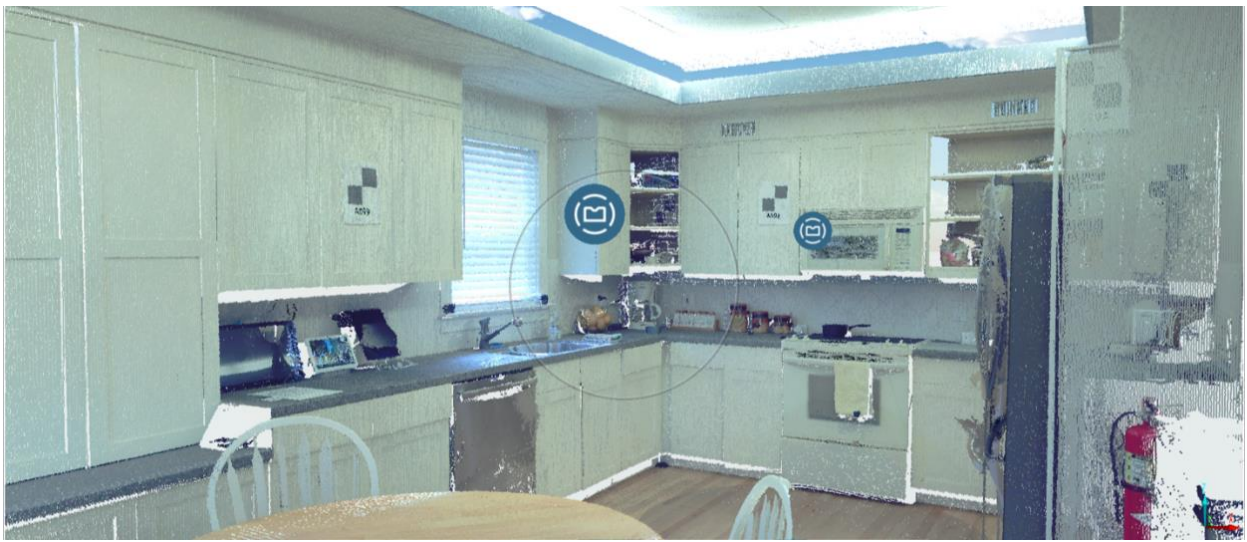

**Figure 2.** Raw point cloud data in Faro SCENE software showing checkerboard registration targets (e.g., A092, A093) placed on cabinetry. These targets enabled precise alignment of the two scan positions.

## 2.3 The Virtual Environment Development Pipeline

The translation of the physical environment into a high-fidelity virtual simulation followed a four-step methodological protocol. This workflow was designed to balance sub-millimeter geometric accuracy with the strict performance budget required for real-time rendering. Specifically, the pipeline targeted a stable 90 Hz frame rate to mitigate cybersickness in older adult participants.

### *Step 1: Point Cloud Processing and Registration*

Raw scan data was imported into Faro SCENE software (Version 2023.1) for initial data hygiene and registration. The primary objective was to register and unify the two individual scans into a single, cohesive point cloud. As illustrated in Figure 2, this was achieved using a target-based registration algorithm that triangulated the spatial coordinates of the checkerboard references placed during scanning. The registration process yielded a mean point error of 1.2 mm, a metric well within the acceptable tolerance for architectural reconstruction (Remondino, 2011). Following registration, a "Stray Point Filter" was applied to eliminate noise artifacts caused by airborne particulates or edge diffraction. A critical manual verification phase was then conducted to address refractive artifacts ("ghosting"). These errors were particularly prevalent on specular surfaces, such as the microwave's glass door and the glazed windowpanes, where the laser beam was either scattered or refracted. These erroneous points were manually excised to prevent geometric distortions in the downstream reference model.

### *Step 2: Data Verification and Format Conversion*

The cleaned point cloud was exported as an unstructured .e57 file, a vendor-neutral format, and imported into Autodesk ReCap Pro for intermediate Quality Assurance (QA). This stage served as a critical interoperability layer between the scientific scanning software (Faro SCENE) and the downstream design tools. As illustrated in Figure 3, ReCap's "RealView" mode was utilized to validate the spatial congruency between the RGB photographic overlay and the depth geometry. This verification confirmed that the texture maps were correctly oriented, ensuring that the visual reference for the modeling phase was dimensionally accurate. Once verified, the cloud underwent volume segmentation to isolate the kitchen boundaries, removing extraneous data from the surrounding hallway and living room. The finalized dataset was then exported as a unified .rcp (Reality Capture Project) file, ready for direct integration into the modeling environment.

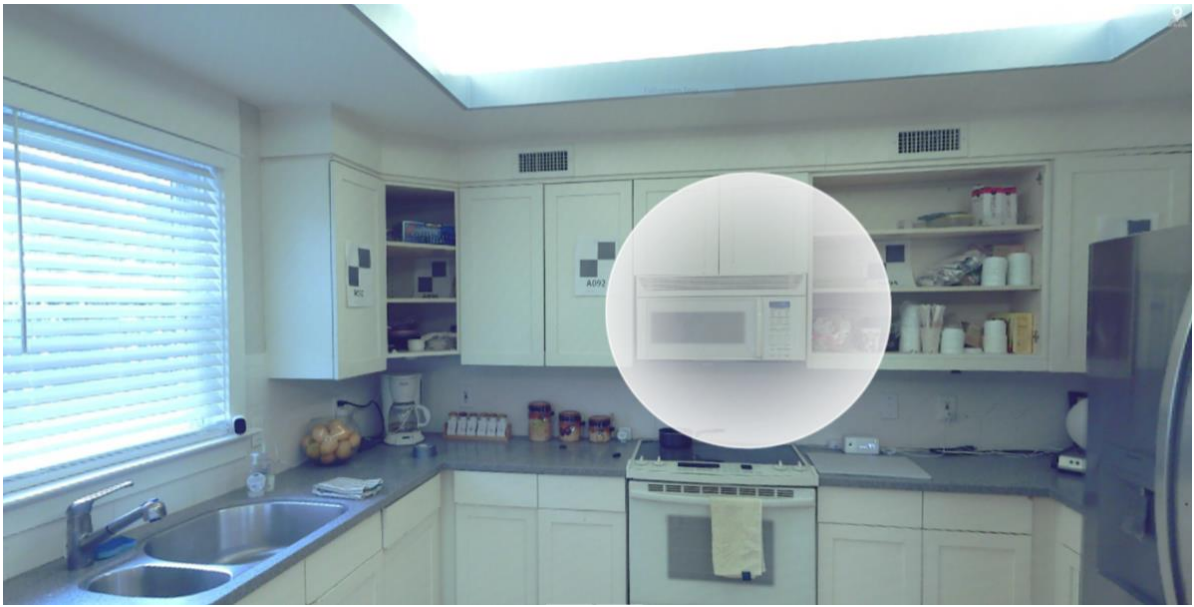

**Figure 3.** Data verification in Autodesk ReCap Pro. RealView inspection confirmed accurate mapping of scanner color data to depth geometry prior to export.

## *Step 3: Geometric Modeling and Retopology*

For the modeling phase, the study chose SketchUp Pro solely for its effectiveness in orthogonal architectural reconstruction. To avoid the stochastic, high-polygon artifacts typical of automated meshing algorithms, which are frequently incompatible with real-time rendering engines, a manual retopology strategy was implemented (Hichri et al., 2013). The unified .rcp point cloud was imported into SketchUp as a volumetric reference layer. The environment was subsequently reconstructed by generating clean, low-polygon geometry directly over the scan data. This methodology ensured that structural elements (walls, countertops, and cabinetry) adhered to the physical dimensions of the Aware Home to within millimeter-grade fidelity, while simultaneously optimizing the mesh topology for high-frame-rate VR performance. This phase also facilitated the precise modeling of the two kitchen configurations under investigation: Condition A (Closed Cabinetry) and Condition B (Open Shelving), using the scan data as a rigid spatial constraint. This approach ensured that both conditions occupied identical spatial footprints, effectively isolating visual accessibility as the sole independent variable by eliminating geometric confounders.

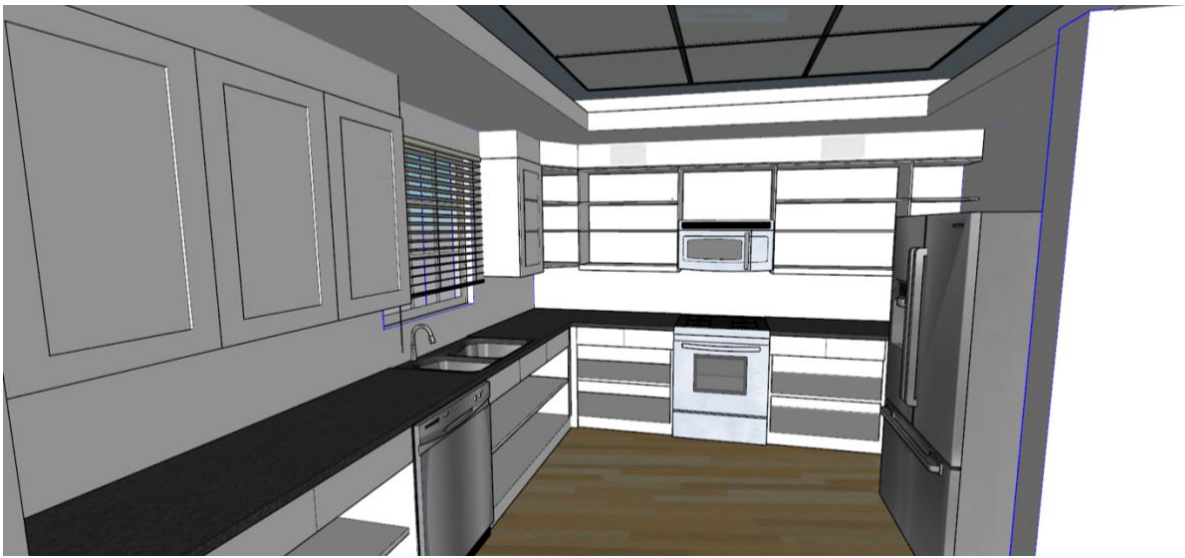

**Figure 4.** Geometric modeling phase in SketchUp. The point cloud serves as a volumetric reference, enabling

precise construction of clean geometry (solid surfaces). The model shows the development of the Open Shelving condition based on the spatial constraints of the original cabinetry.

## *Step 4: Engine Integration and Simulation*

The optimized 3D assets were migrated into Unreal Engine 5.3.2 (UE5) via the Datasmith pipeline. Datasmith was prioritized over standard FBX export protocols because it preserves the semantic hierarchy and metadata of the SketchUp model. This data retention is essential for efficient iteration, allowing design variables such as cabinet types to be swapped without losing their coordinate references. To achieve photorealistic fidelity, the study utilized UE5's Lumen dynamic global illumination system. Unlike traditional baked lighting, Lumen calculates diffuse interreflection in real time, enabling the simulation to accurately replicate the complex light transport in the physical kitchen. Specifically, invisible RectLight actors were embedded in the virtual drop ceiling to mimic the soft, indirect illumination of the physical cove lights. This lighting is strictly governed by visual accessibility principles; as noted by the IES (2020), older adults require higher contrast levels and minimized glare for effective object recognition and spatial orientation. Finally, interaction mechanics were implemented using the *OpenXR* standard to ensure hardware agnosticism. Collision primitives (capsules) were generated for all counters and appliances to enforce kinematic boundaries. This prevents participants from passing through solid geometry, thereby maintaining the simulation's physical laws and preserving the Plausibility Illusion. It should be noted that the visual output experienced by participants through the HTC VIVE Pro HMD differs perceptually from engine viewport screenshots due to HMD tone mapping, lens distortion correction, and AMOLED display characteristics; the viewport captures presented in Figure 5 are therefore illustrative of scene composition rather than a direct representation of the participant experience.

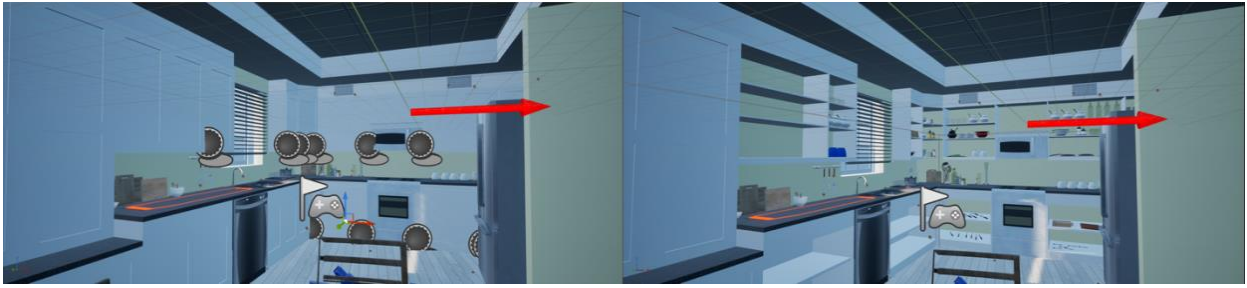

**Figure 5.** Unreal Engine 5.3.2 viewport captures of Condition A (Closed Cabinets) and Condition B (Open Shelving). Lumen Global Illumination was configured to match the luminance characteristics of the physical cove lighting; apparent tonal differences from Figure 1 reflect engine viewport rendering characteristics rather than lighting model deviation and are not representative of the participant experience through the HMD. Arrows indicate the Widget Interaction Component direction vectors attached to each hand controller, visible in the editor viewport by default.

## 2.4 Participants

A within-subjects VR study was conducted with 17 community-dwelling older adults (7 with MCI confirmed by the Montreal Cognitive Assessment (MoCA) score ≤ 25, 10 without MCI; mean age 77.5 years [SD = 6.3]; age data available for 13 of 17 participants). The study employed a counterbalanced design in which each participant experienced both kitchen conditions (Closed Cabinetry and Open Shelving). Condition order was determined using block randomization with block sizes of 4 (Efird, 2011) to ensure balanced allocation and that Setting 1 could be either condition with equal probability across participants.

## 2.5 Materials

The Simulator Sickness Questionnaire (SSQ; Kennedy et al., 1993) was administered at three time points: immediately before the first condition (baseline), immediately after Setting 1, and immediately after Setting 2. The NASA Task Load Index (NASA-TLX; Hart & Staveland, 1988) was administered immediately after each condition (Setting 1 and Setting 2). The final simulation was administered via the HTC VIVE Pro Head-Mounted Display (HMD). This specific HMD was selected for its high-resolution architecture and precise tracking capabilities, which are technical prerequisites for maintaining immersion fidelity and minimizing cybersickness in older adult participants. The HMD features dual AMOLED screens with a combined resolution of 2880 × 1600 pixels (1440 × 1600 pixels per eye). The displays operate at a refresh rate of 90 Hz, a critical threshold for ensuring fluid motion and vestibular comfort during rapid head rotation (HTC Corporation, 2018). For spatial positioning, the system utilizes SteamVR Tracking 2.0 with dedicated base stations. This technology enables sub-millimeter spatial precision, translating the user's physical movements within the kitchen. Additionally, the headset includes integrated Hi-Res-certified headphones that deliver spatialized auditory feedback, further reinforcing the Plausibility Illusion.

## 2.6 Procedure

After providing informed consent and completing baseline demographics and cognitive screening, participants donned the HTC VIVE Pro HMD and performed a standardized IADL task, simulated kitchen activity, mainly object retrieval from cabinets/shelves, and placement on the counter, in each condition. Short breaks were provided between conditions to minimize fatigue. All procedures were approved by the college's Institutional Review Board.

## 2.7 Hardware & Software Specifications

To guarantee the replicability of this methodological pipeline and establish a benchmark for performance optimization, the specific hardware and software configurations employed throughout the development and deployment phases are detailed. A comprehensive summary of these specifications is provided in Table 1.

### 2.7.1 Environmental Scanning Hardware

The digitization of the physical environment was executed using the FARO Focus S70 laser scanner. This device was selected for its metrology-grade measurement capabilities, which establish the ground truth for indoor architectural documentation. The scanner features a systematic ranging error of ±1mm and a range noise of 0.3mm at a 10m distance on 90% reflective surfaces (Faro Technologies, 2021). The device captures a spherical field of view (300° vertical × 360° horizontal), ensuring comprehensive coverage of the interior envelope, including ceiling plenums and floor transitions. Furthermore, the scanner integrates an HDR camera with exposure bracketing to capture up to 165 megapixels of color data. This radiometric capability is critical for generating photorealistic textures that match the geometric depth data. The system also supports in-situ registration via the proprietary software interface, enabling real-time quality control of scan density and overlap during scanning.

### 2.7.2 Development Workstation

The processing of massive point cloud datasets and the real-time rendering of the Unreal Engine simulation necessitated a high-performance computing (HPC) environment. All development was conducted on an Alienware Aurora R12 workstation. The system architecture incorporated an 11th Gen Intel® Core™ i9-11900F processor, clocked at 2.50GHz. This multi-core CPU provided the necessary parallel processing power for registering large-scale point cloud datasets. To mitigate memory bottlenecks during the manipulation of unoptimized high-poly meshes, the system was equipped with 32.0 GB of DDR4 memory running at 3200 MHz. Graphics processing was offloaded to an NVIDIA GeForce RTX 3080 with 10 GB of VRAM. This GPU was critical for supporting hardware-accelerated Ray Tracing, a prerequisite for the Lumen global illumination system used in Unreal Engine 5. For storage, the system used a hybrid configuration totaling 1.84 TB, comprising a 954 GB NVMe SSD (Micron 2300) for high-speed software execution and data caching, and a 932 GB HDD for archiving raw scan files.

**Table 1.** *Hardware and Software Specifications for the Environmental Development Pipeline*

| Category | Component | Specification | Methodological Role |
|---|---|---|---|
| **Scanning** | **Scanner** | Faro Focus S70 Laser Scanner | Volumetric Spatial Capture |
| | Resolution | 1/5 Resolution (approx. 44 MP) | High-Fidelity Texture Mapping |
| | Accuracy | ±1mm ranging error @ 10m | Geometric Ground Truth |
| **Processing** | **Workstation** | Alienware Aurora R12 | Computational Processing |
| | CPU | Intel Core i9-11900F @ 2.50GHz | Parallel Registration Algorithm |
| | GPU | NVIDIA GeForce RTX 3080 (10GB) | Lumen Ray Tracing Acceleration |
| | RAM | 32 GB DDR4 @ 3200 MHz | High-Poly Mesh Manipulation |
| **Deployment** | **VR Headset** | HTC VIVE Pro | Experimental Interface |

|  | Display | Dual AMOLED 3.5" (1440 x 1600/eye) | Retinal Resolution |
|  | Refresh Rate | 90 Hz | Vestibular Comfort (Cybersickness Mitigation) |
|  | Tracking | SteamVR Tracking 2.0 | 1:1 Kinematic Fidelity |

## 3. Results

### 3.1 Visual Fidelity and Ecological Validity

The overarching goal of this methodological pipeline was to generate a photorealistic simulation capable of eliciting a sense of presence comparable to that of the physical environment. To assess the environmental congruence of the simulation, a direct, side-by-side comparison was conducted between the source photography and the final rendered output (captured as viewport screenshots within the Unreal Engine 5 editor prior to headset deployment). As illustrated in Figures 1 and 5, the developed pipeline successfully retained the spatial integrity of the Aware Home kitchen. The application of the Lumen global illumination system accurately replicated the soft shadow gradients cast by the cove lighting. While the captured Figure 5 render may appear cooler in tone than the physical setting due to display calibration, the Lumen system was configured to match the measured luminance characteristics of the cove fixtures per IES (2020) guidelines. These lighting cues are critical for depth perception and spatial orientation, particularly for older adults who rely on contrast for navigation. Furthermore, the "retopology" modeling strategy preserved the exact architectural proportions of the cabinetry. This ensured that virtual reach distances matched the physical constraints of the real world with millimeter-grade fidelity. This structural isomorphism is a prerequisite for the ecological validity of the study. It guarantees that any biomechanical data collected in VR, such as reaching height or bending angle, is representative of real-world behavior rather than an artifact of an incorrect virtual scale.

### 3.2 System Performance Optimization

High-fidelity scan data carries a significant computational overhead. If unoptimized, this geometric density can result in rendering latency that precipitates vection-induced "cybersickness," a phenomenon to which older adults are particularly susceptible (Stanney et al., 2020). Consequently, the technical success of this pipeline was quantified by its capacity to sustain a stable frame rate within the strict latency budget required for VR comfort. Following the optimization phase in SketchUp, where back-face culling and mesh topology simplification were rigorously applied, the final scene asset was reduced to approximately 450,000 polygons. On the deployment workstation equipped with the NVIDIA RTX 3080, the simulation sustained a mean frame rate of 90.4 FPS (Frames Per Second) with an average frame time of 11.1 ms, well within the 11.1 ms frame time budget required to maintain a native 90 Hz refresh rate. This data confirms that the pipeline produces a simulation that is clinically viable for use with sensitive populations, specifically optimizing for the vestibular comfort of older adults with MCI.

### 3.3 Participant Comfort and Functional Feasibility Validation

To confirm the pipeline's tolerability for the target population and its sensitivity to the experimental manipulation, SSQ and NASA-TLX scores were examined across both conditions.

As shown in Table 2, SSQ total scores exhibited no statistically significant increase from baseline to the closed-cabinet condition (Wilcoxon signed-rank, p = .279) or the open-shelving condition (p = .182), nor between conditions (p = .515). Median scores remained stable or decreased slightly, with no sessions terminated due to discomfort. These findings confirm that the sub-millimeter geometric fidelity, Lumen global illumination, and sustained 90 Hz performance successfully mitigated cybersickness risk even among participants with MCI.

NASA-TLX workload scores were significantly lower in the open-shelving condition (M = 6.61, SD = 3.49) than in the closed-cabinet condition (M = 8.32, SD = 3.16; Wilcoxon signed-rank, p = .019, effect size r = .41), demonstrating that the simulation preserved sensitivity to the visual-accessibility manipulation without introducing confounding workload artifacts. This pattern aligns with earlier evidence showing reduced cognitive load and improved task engagement in open-shelving environments for older adults with MCI (Bilau et al., 2025).

**Table 2.** *SSQ and NASA-TLX scores across conditions (N = 17)*

| Measure | Baseline | Closed Cabinets | Open Shelving |
|---|---|---|---|
| **SSQ Total – Median (IQR)** | 56.7 (0–99.7) | 52.1 (0–172.8) | 28.3 (0–160.8) |
| **NASA-TLX – M (SD)** | – | 8.32 (3.16) | 6.61 (3.49) |

*Note.* SSQ scores are reported as medians (and IQR) due to positive skew from a small number of higher-scoring MCI participants. All scores remained within tolerable ranges reported in prior VR aging studies.

# 4. Discussion

## 4.1 Bridging the Gap Between Engineering and Behavioral Science

The primary contribution of this study is the codification of a Reality-to-VR pipeline that bridges the metrological precision of engineering surveys with the ecological requirements of behavioral science. Prior research has frequently relied on manual 3D modeling to generate experimental environments (Brookes et al., 2020). While sufficient for abstract cognitive tasks, manual modeling often yields idealized geometries devoid of the stochastic irregularities and environmental entropy inherent to real-world domestic settings. By establishing TLS as the geometric ground truth, this protocol ensures that the virtual environment captures the exact spatial affordances of the physical home. For aging research, this distinction is critical. Older adults rely heavily on familiar environmental cues for sensorimotor integration and task performance (Davis et al., 2009). The sub-millimeter accuracy achieved by the Faro scanner ensures that biomechanical interactions, such as the reach distance to a shelf or the trunk flexion required for an oven, possess high kinematic validity. This rigorous correspondence strengthens the plausibility that virtual findings will transfer to real-world applications, a hypothesis that future comparative studies, contrasting scan-based and synthetic environments on matched behavioral tasks, are now positioned to test directly using this pipeline as a benchmark. Furthermore, this digital workflow introduces a level of experimental agility impossible to achieve in a physical setting. In contrast to physical renovations, which are cost-prohibitive and static, the virtual environment permits the instantaneous manipulation of architectural variables. This capability facilitates rigorous within-subjects experimental designs, allowing participants to evaluate multiple spatial configurations, such as open shelving vs. closed cabinet, in rapid succession without the logistical constraints of a real-world remodel. Participant validation with 17 older adults (7 with MCI) further substantiated the pipeline's readiness for the target population: SSQ scores showed no statistically significant increase from baseline across both conditions (median scores remained low and stable), confirming effective cybersickness mitigation even among individuals with MCI. NASA-TLX scores were significantly lower in the open-shelving condition ($p = .019$, $r = .41$), demonstrating that the simulation preserved sensitivity to the visual-accessibility manipulation without introducing confounding workload artifacts.

## 4.2 Technical Challenges and Recommendations

While the pipeline successfully generated a high-fidelity photorealistic simulation, several technical bottlenecks were identified that future researchers must navigate to ensure reproducibility.

***The Data Density Constraint:*** The first major hurdle was the sheer volume of raw point cloud data. Processing clouds with millions of points posed significant computational overhead, often causing instability on standard hardware. This necessitated the rigorous optimization strategies described in the methodology to maintain software responsiveness.

***Specular Reflection Mitigation:*** A second significant challenge was managing "ghosting" artifacts during point cloud processing in Faro SCENE. Highly reflective surfaces, specifically the glass microwave door and the stainless-steel refrigerator, caused laser beam scattering, resulting in erroneous data points. Cleaning these artifacts required substantial manual intervention, constituting approximately 30% of the total post-processing time. To mitigate this in future studies, researchers are recommended to physically remove or mask specular objects in the room prior to scanning. This preemptive step can significantly reduce post-processing labor.

***Semantic Segmentation and Engine Integration:*** The final critical challenge involved the transition from the point cloud to the game engine. Directly importing raw point clouds into Unreal Engine incurs significant performance overhead and lacks the semantic segmentation required for interaction, such as identifying a door vs. a wall. The intermediate retopology phase in SketchUp was therefore essential. Although time-intensive, manually tracing the geometry resulted in a dramatic reduction in polygon count without sacrificing perceived visual fidelity. For this transition, using the Datasmith exporter is specifically recommended. Unlike standard file formats such as FBX, Datasmith preserves the complex object hierarchy and metadata from the modeling software, which is essential for defining interactive elements and logic within the game engine.

## 4.3 Limitations

Despite its strengths, this workflow presents specific limitations that must be acknowledged. First, the High-Performance Computing (HPC) prerequisites are substantial. Processing raw scan data and rendering real-time global illumination requires a workstation with significant memory bandwidth (32 GB+ RAM) and a high-end GPU (RTX 3080 or equivalent) to support the Lumen system. This computational demand may present an interdisciplinary resource barrier for behavioral science departments lacking dedicated engineering computing infrastructure. Second, the environment exhibits temporal rigidity. While the lighting simulation is dynamic, the scanning process captures a frozen moment. Consequently, any interactive elements, such as opening drawers or repositioning appliances, cannot be simply "turned on." They must be semantically separated from the static mesh and kinematically articulated (rigged). This requirement adds significant complexity to the development cycle compared to fully procedural or synthetic environments.

## 5. Conclusion

This paper has codified a reproducible, end-to-end protocol for converting existing built environments into photorealistic VR simulations for behavioral research. By synthesizing TLS with real-time global illumination technologies, the feasibility of creating photorealistic environmental simulations that reconcile the rigorous dimensional fidelity demanded by engineering standards with the high plausibility requirements of psychological immersion has been demonstrated. Participant validation confirmed minimal cybersickness risk (SSQ) and preserved sensitivity to the experimental design variable (NASA-TLX) in the target older-adult population, including individuals with MCI. For the domain of Cognitive Aging in Place, this methodology offers a validated alternative to reductionist simulation. It facilitates a transition from abstract testing to ecologically valid assessment, enabling IADLs to be observed in environments that preserve the stochastic complexity of the real world. As reality capture hardware and real-time rendering engines continue to converge, this "Reality-to-VR" workflow establishes a nascent gold standard for rigorous environmental design research.